\documentclass[12pt,a4papper]{article}

\usepackage{amssymb}

\begin{document}

\title{Pioneer acceleration and variation of light speed: experimental situation }

\author{Antonio F. Ra\~nada\\Facultad de F\'{\i}sica, Universidad Complutense\\ E-28040 Madrid, Spain}

\date{2 January 2004}

\maketitle

\thispagestyle{empty}

\begin{abstract}

The situation with respect to the experiments is presented
of a recently proposed model that gives an explanation of the
Pioneer anomalous acceleration $a_{\rm P}$. The model is based on an
idea already discovered by Einstein  in 1907: the light speed depends
on the gravitational potential $\Phi$, so that it is larger the higher
if $\Phi$. The potential due to all the mass and energy in the universe
increases in time because of its expansion, which has the consequence that light must be slowly accelerating.
Moreover it turns out
that the observational effects of a universal adiabatic acceleration of light $a_\ell
=a_{\rm P}$ and of an extra acceleration towards the Sun $a_{\rm P}$ of a spaceship would be the same:
a blue shift increasing linearly in time, precisely what was observed.
The phenomenon would be due to a cosmological acceleration of the proper time of bodies with respect to the coordinate time. It is shown that it agrees with the
experimental tests of special relativity and the weak equivalence
principle if the cosmological variation of the fine structure
constant is zero or very small, as it seems now.

\bigskip


\end{abstract}

\newpage
\pagestyle{empty}

\tableofcontents

\newpage

\setcounter{page}{1}

\pagestyle{myheadings}

\section{Introduction}
The purpose of this paper is to discuss the situation with respect
to the observations of the recent proposal that there could be an
adiabatic universal acceleration of light of the order of $a_\ell
\approx 10^{-9}\mbox{m/s}^2$ \cite{Ran03a,Ran03b,Ran03c}. Such an
acceleration  would give an explanation for one of the most
intriguing riddles in today's cosmology: the anomalous and
unmodelled acceleration observed in the Pioneer 10/11, Galileo and
Ulysses spacecrafts. It would be explained then as an effect of
the adiabatic attenuation of the quantum vacuum, which would be cause
the proper time of bodies to accelerate  with respect to the
coordinate time. Strange as it
might seem, this value is so small that it is plausible that it
could have not been detected, in spite of the enormous number of
optical phenomena which are studied and analyzed daily.

 In October 1998, Anderson {\it et al}
reported a tiny but significant anomaly in the motion of the
Pioneer 10/11 spaceships: the solar attraction seems to be
slightly larger than what predicted by standard gravity \cite{And98}.
Their analysis of the data from the two missions found in the
motion of both spacecrafts  an extra unmodelled constant
acceleration towards the Sun, equal to $a_{\rm P}\simeq 8.5\times
10^{-10}$ m/s$^2$. The data of the Galileo and Ulysses spaceships
showed the same effect. Surprisingly, no similar extra acceleration
was found in the case of the planets, as it would be required by
the equivalence principle if the effect were due to gravitational
forces (of dark matter for instance). Anderson {\em et al} stated
``it is interesting to speculate on the possibility that the
origin of the anomalous signal is new physics." In spite of a
thorough search, they could not find as yet any reason for that
extra acceleration (see \cite{And02} for a detailed review of the
problem and of the observational techniques involved).

It was suggested in 2003 by this author that this effect could be
explained if the light is increasing its speed with an adiabatic
acceleration $a_\ell$ equal to $a_{\rm P}$
\cite{Ran03a,Ran03b,Ran03c}. More precisely, a mechanism was
proposed in these papers which must produce such an acceleration
as a consequence of the combined effect on the quantum vacuum of
(i) the fourth Heisenberg relation and (ii) the expansion of the
universe, the treatment being Newtonian and phenomenological. It
follows from this mechanism that the electric and magnetic constants
of empty space, $\epsilon _0$ and $\mu _0$, are decreasing adiabatically
(and its optical density as well), the
light speed increasing consequently. Curiously enough, as shown in
\cite{Ran04}, it turns out that a general relativistic argument
predicts the existence of such an acceleration if the expansion of
the universe is taken into account, its value being close, at
least, to $a_\ell \simeq H_0c\approx 0.8a_{\rm P}$, $H_0$ being
the present value of the Hubble parameter. Furthermore, an
adiabatic acceleration $a_\ell$ would imply a blue shift in the
radio signal of the spacecrafts, which increases linearly in time
such that $\dot{\nu} =\nu a_{\rm P}/c$, precisely what was
observed by Anderson {\it et al}. The possibility must be studied
that the ships did not accelerate but followed instead the laws of
standard gravity, the blue shift not being due to their motion but
to the acceleration of light. The model based on that mechanism
will be called here ``this model" or ``the attenuating vacuum
model".

However and understandably, the first reaction of many physicists
would probably be that the light speed is too much well known to
admit such an acceleration without violating some of the
experimental bounds imposed by the tests of special relativity and
the equivalence principle, or that it would have been detected in
interferometry, optical communication and similar effects.

 \section{The mechanism}
In this section a terse summary of the attenuating vacuum model is
given, in the frame of a Newtonian approximation, although with
the inclusion of the rest energy of the particles.
 As is explained in previous work \cite{Ran03a,Ran03b,Ran03c}, it
 can be argued that the quantum vacuum can be compared to an
 inhomogeneous optical medium such that its optical density depends on
 the gravitational potential. This dependance would be important since the
quantum vacuum fixes some important quantities through
renormalization processes, as is the case of the electron charge
$e$ (and eventually of the monopole charge), the light speed and
the fine structure constant. A coupling between its density and
the gravitational field would entail a space and time
inhomogeneity across the universe of such quantities as $e$, $c$
and $\alpha$. To understand why, note that the empty space can be
considered as a sea of virtual pairs and other particles which pop
up and disappear continuously. On the average, a pair created with
energy $E$ (including rest energy, kinetic energy and
electromagnetic energy) lives during a time $\tau \approx \hbar
/E$ according to the fourth Heisenberg relation. If there is a
Newtonian gravitational potential $\Phi$, the pair pops up with an
extra energy $E\Phi /c^2$, so that its lifetime must
\begin{equation}
\tau _\Phi =\hbar/(E+E\Phi /c^2)= \tau _0/(1+\Phi c^2 ).
\label{10}
\end{equation}
 As a clear if unexpected
consequence, the number density of pairs $\cal N$ would depend on
the gravitational potential $\Phi$ as
\begin{equation}
{\cal N}_\Phi ={\cal N}_0/ (1+\Phi /c^2). \label{20}
\end{equation}
 In other words, the optical density of the quantum
vacuum would vary in space and time, because it must have there an
extra number density of pairs depending on $\Phi$. In this model,
the lower (more negative) is $\Phi$, the denser is the quantum
vacuum, and conversely the higher (less negative) is $\Phi$, the
thinner is the quantum vacuum.
 Note that this variation
of the  virtual pairs lifetime in a gravitational field is not an
{\em ad hoc} hypothesis, but an compelling consequence of the
fourth Heisenberg relation. Note also that the virtual particles
considered here are not created by gravity: they are just the
usual zero point  particles that fill the space everywhere,
according to elementary quantum physics: they live a bit longer
(or shorter), that is all.

It is unavoidable here to take $\Phi$ as the gravitational
potential created by all the universe, so that this model has
something in common with the Mach principle, since some important
properties of the bodies are determined by their interaction with
all the matter and energy in the visible universe.
Indeed the gravitational potential energy of a body is defined as
the energy required to bring it from the infinity to its actual
position, without changing its kinetic energy or any other
nongravitational energy. It could be argued that a body can not be
brought actually from the infinity, since this is farther away
than the horizon of the visible universe. However, it is clear
that less energy is needed to create a virtual pair if $\Phi$ is
negative than if it is zero. Therefore, the variation of the
lifetime of the virtual pairs is due in this model to the
gravitational interaction of the virtual pairs of the quantum
vacuum with all the matter and radiation in the universe, and also
with all the quantum vacuum itself.

As a result of
the previous arguments some physical constants must vary in space
and time because they are renormalized in a way that depends on
$\Phi$. Any one, say $f$, can be expressed, at first order in the
potential, either as $f({\bf r}, t)=f_\infty [1+ \sigma \,\Phi
/c^2]$ or as $f({\bf r},t)=f_{\rm R} [1+ \sigma ^\prime (\Phi
-\Phi _{\rm R})/c^2]$, where $\Phi =\Phi ({\bf r},t)$, the
subindexes $\infty$ and ${\rm R}$ indicate value at zero potential
and at a reference terrestrial laboratory R, respectively, and
$\sigma$, $\sigma^\prime$ are two coefficients. We will consider
only cases in which $(\Phi -\Phi _{\rm R})/c^2$ is small.

 In particular, the relative
permittivity and permeability of empty space must depend on the
gravitational potential, their expressions at first order being
\begin{equation}
\epsilon _{\rm r}({\bf r},t)=1-\beta [\Phi ({\bf r},t) -\Phi _{\rm
R}
 ]/c^2,\;\;\;\; \mu _{\rm r}({\bf r},t)=1-\gamma [\Phi ({\bf
r},t)-\Phi _{\rm R} ]/c^2, \label{30}
\end{equation}
where $\beta$ and $\gamma$ are certain coefficients, which must be
positive since the quantum vacuum is dielectric but paramagnetic
(its effect on the magnetic field is due to the magnetic moments
of the virtual pairs). For later convenience, we will used instead
the combinations $\chi =(\beta +\gamma)/2$, $\xi =(3\beta
-\gamma)/2$ [so that $\beta =(\chi +\xi )/2,\, \gamma
=(3\chi-\xi)/2$]. Clearly, $\epsilon _{\rm r}=1,\,\mu_{\rm r}=1$
 at the reference laboratory at
Earth.

This implies that the electron charge, the light velocity
and the fine structure constant do vary in the universe, their
values at a generic spacetime point $({\bf r},t)$ being expressed,
at first order, as
\begin{eqnarray}
e({\bf r},t)&=& e[1+ {\chi +\xi \over 2} \, {\Phi -\Phi _{\rm R}\over c^2}],\label{40}\\
c({\bf r},t) &=& c[1+\chi\, {\Phi -\Phi _{\rm R} \over c^2}],\label{50}\\
\alpha ({\bf r},t) &=& \alpha [1+ \xi \, {\Phi -\Phi _{\rm R}\over
c^2}],. \label{60}
\end{eqnarray}
Here and along this paper $e,\, c, \, \alpha$ in an equation (without spacetime dependence) will be
the values now at Earth ({\it i. e.} the constant in the
tables). When this need to be emphasized, they will be denoted $e_0, c_0, \alpha _0$, unless otherwise specified.
When they are considered to be variables, they will be written with space and/or time arguments.
It was shown in \cite{Ran03a} that eq. (\ref{60}) agrees with the
observed time evolution of the fine structure constant
\cite{Web01} if $\xi =(3\beta -\gamma)/2\approx 1.3 \times
10^{-5}$. If the variation of $\alpha$ has been overestimated, as
indicated by some recent data \cite{Sri04}, then $\xi$ would be
smaller.  If $\alpha$ does not vary, then $\xi=0$ (and $\gamma
=3\beta$). On the other the best value for $\chi$ turns out to be
$\chi =1$, as was shown in \cite{Ran03c} and will be discussed later.

 Note that the
variation of the potential between two spacetime points can be
written as the sum of a space
and a time variation $\Delta \Phi =\Delta _{\rm s}\Phi +\Delta _{\rm t}\Phi$,
where  $\Delta _{\rm s} \Phi = \Phi ({\bf r}_2,t_2)- \Phi ({\bf r}_1, t_2)$ and
$\Delta _{\rm t} \Phi = \Phi ({\bf r}_1,t_2)- \Phi ({\bf r}_1, t_1)$.
In this paper the point 1 will be the laboratory R at present time $t_0$, unless otherwise specified.
The two variations  have different effects, as was
shown in \cite{Ran03b,Ran03c}. This is important since one or
the other of the two variations can be neglected in some
interesting cases. In the experiments considered here, the time
variation can be neglected, while in the observations leading to the discovery
of the Pioneer acceleration, it is dominant.

(i)  The spatial variation of $\Phi$ causes the light to behave as in an ordinary
inhomogeneous optical medium, in such a way that the frequency remains constant
during the propagation, while the wavelength and the light
velocity change according to the value of a refractive index $n$, which depends on
$\Phi$ as \begin{equation} n =\{1+\chi [\Phi -\Phi _{\rm
R}]/c^2\}^{-1}\label{70} ,\end{equation}
This space variation can be either positive or
negative according to how much matter is around. It follows from (\ref{50}) that $c$ is
smaller where $\Phi$ is more negative (or less positive), {\it
i.e.} it decreases when approaching massive objects. In the case of the
Sun surface, for instance, $(\Phi _{\rm S}-\Phi _{\rm R})/c^2\simeq
-2.1\times 10^{-6}$,  with $\Phi _{\rm S}$ being the potential there
(and $n= 1-2.1\times 10^{-6}$, so that light speed
is about 2 ppm smaller than here).
If, however, we take Jupiter instead of the Sun, then $\Delta \Phi >0$ and $n>1$.

(ii)  The time variation of $\Phi$, as was shown in \cite{Ran03b,Ran03c} and will be
seen here, is dominated by a secular component due to the expansion
of the universe. This is because the progressive separation of the
galaxies implies a monotonous increase of $\Phi$. As a
consequence, $\epsilon _{\rm r}$, $\mu _{\rm r}$ decrease slowly
 what must cause an adiabatic acceleration of
light. It can be shown and will be seen later that  an adiabatic
acceleration of the light speed causes an linear increase of the
frequency such that $\dot{\nu} =\nu a_{\rm P}/c$, the wavelength
remaining constant. This variation of $c$ is the most important prediction
 of the model, although it is negligible in short time
terrestrial laboratory experiments.

The potential at the terrestrial laboratory $R$ can be written,
with good approximation, as $\Phi =\Phi _{\rm inh}(R)+\Phi_{\rm
av}(t)$. The first term $\Phi _{\rm inh}(R)$ is the part due to
the near local inhomogeneities (the Earth, the Solar System and
the Milky Way). It can be taken as constant since these objects
are not expanding. The second $\Phi_{\rm av}(t)$ is the space
averaged potential due to all the mass and energy in the universe,
assuming that they are uniformly distributed. It depends on time
because of the expansion. The former has a non vanishing gradient
but is small, the latter is space independent (it has zero
gradient)
 but is time dependent and much larger. The
 value of $\Phi _{\rm
inh}/c_0^2$ at $R$ is the sum of the effects of the Earth, the Sun
and the Milky Way, which are about $-7\times 10^{-10}$, $-10^{-8}$
and $-6\times 10^{-7}$, respectively, which have much smaller
absolute values than $\Phi _{\rm av}$, which of the order of
$-10^{-1}$ as will be seen below. Moreover $\Phi _{\rm inh}$
appear in the two terms of the difference of potential in eq.
(\ref{30}), so that, being time independent, it cancels out. In
fact, that difference is $[\Phi_{\rm av}(t)+ \Phi _{\rm
inh}(R)-(\Phi_{\rm av}(t_0)+ \Phi _{\rm inh}(R)]=[\Phi _{\rm
av}(t)-\Phi _{\rm av}(t_0)]$. This means that eq. (\ref{30}) can
be written at R as
\begin{equation}
\epsilon _{\rm r}=1-\beta [\Phi_{\rm av}(t) -\Phi _{\rm av}
(t_0)]/c^2,\;\;\;\; \mu _{\rm r}=1-\gamma [\Phi_{\rm av}(t) -\Phi
_{\rm av}(t_0)]/c^2, \label{80}
\end{equation}
where $\Phi _{\rm av}(t)$ is the space averaged gravitational
potential of all the universe at time $t$ and $t_0$ is the present
time ({\em i.e.} the age of the universe).

Let $\Phi _0$ be the gravitational potential produced by the
critical density distributed up to the distance of $R_U$ ($\Phi
_0=-\int _0^{R_U} G\rho _{\rm cr}4\pi r{\rm d}r\simeq -0.3c^2$ if
$R_U\approx 3,000$ Mpc)
  and let $\Omega _M,\,\Omega _\Lambda$ be the corresponding
present time relative densities of matter (ordinary plus dark) and
dark energy corresponding to the cosmological constant $\Lambda$.
Because of the expansion of the universe, the gravitational
potentials due to matter and dark energy equivalent to the
cosmological constant vary in time as the inverse of the scale
factor $R(t)$ and as its square $R^2(t)$, respectively. It turns
out therefore that
\begin{equation}
\Phi_{\rm av}(t)-\Phi_{\rm av}(t_0)=\Phi _0F(t),\;\;\;\mbox{ with
}\;\;
  F(t)=\Omega
_M[1/R(t)-1]-2\Omega_\Lambda [R^2(t)-1], \label{90}
\end{equation}
where $F(t_0)=0$, $\dot{F}(t_0)= -(1+3\Omega _\Lambda )H_0$. Let
us assume a universe with flat sections $t=\mbox{constant}$ ({\it
i.e.} $k=0$),  with $\Omega _M=0.27$, $\Omega _\Lambda =0.73$ and
Hubble parameter to $H_0=71\mbox{ km}\cdot \mbox{s}^{-1}\cdot
\mbox{Mpc}^{-1}= 2.3\times 10^{-18}\mbox{ s}^{-1}$. To find the
evolution of the average light speed, $\Phi_{\rm av}(t)-\Phi_{\rm
av}(t_0)$ must be substituted for $\Phi ({\rm r},t) -\Phi _{\rm R}$
in (\ref{50}), what gives
\begin{equation}
c(t)=c[1+\chi \, F(t){\Phi _0\over c^2}].
\label{100}\end{equation} The effects of the inhomogeneities of
the Milky Way, the Sun and the Earth  at a terrestrial laboratory
at present time $t_0$ are in fact included here in the numerical
value of $c=c(t_0)$. Taking now the time derivative of this
equation, the present value of the acceleration $a_\ell
=\dot{c}(t_0)$ is found to be
\begin{equation}
a_\ell = a_{\rm t}c,\qquad\mbox{with}\qquad a_{\rm t}=-H_0\chi (1+3\Omega _\Lambda ) \Phi
_0/c^2. \label{110}
\end{equation}
Note that, with $\chi=1$ and the values given before for $\Phi_0$
and $\Omega _\Lambda$, then $a_{\rm t}=2.2\times 10^{-18}\mbox{ s}^{-1}=
 0.96H_0$ and $a_\ell \simeq H_0c= 6.9\times 10^{-10}\mbox{ m/s}^2
 \approx 0.8 a_{\rm P}$. In other words, $a_\ell $
is close to the Pioneer acceleration $a_{\rm P}$. The quantity
$a_{\rm t}$ represents an acceleration of time since $c(t)=c_0(1+a_{\rm t}t)$.
  Note that it plays
the role of an acceleration of the clocks; it will be discussed
in section 5.1. The suggestion that $a_{\rm t}=H_0$ is clear and intriguing. (Remember that
 the best values of these two coefficients were shown to be
$\chi =1$ and $\xi\simeq 10^{-5}$ in order to keep the agreement of the
predicted time variation of $\alpha$ with the observations and to explain
the Pioneer anomaly (see paragraph after eq. (\ref{60})).

\section{The acceleration of light and the blue shift}
It will be shown in this section that {\it an adiabatic
acceleration of light implies a blue shift}, because the frequency
$\nu _0$ of a monochromatic light wave with such an adiabatic
acceleration $a_\ell$ increases so that its time derivative
$\dot{\nu}$ satisfies
\begin{equation}
{\dot{\nu}\over \nu_0} ={a_\ell \over c}. \label{120}
\end{equation}
The consequence is that an adiabatic acceleration of light has the same radio
signature as a blue shift of the emitter, although a peculiar blue
shift with no change of the wavelength ({\em i.e.} all the
increase in velocity is used to increase the frequency).

 Equations (\ref{80}) state that the time derivatives of the permittivity
 $\epsilon =\epsilon _{\rm r}\epsilon _0$ and permeability $\mu =\mu _{\rm r}\mu _0$
  of empty space at present time $t_0$ are equal to
\begin{equation}
\dot{\epsilon} = \epsilon _0\beta {\Phi _0\over c^2}(1+3\Omega
_\Lambda )H_0,\;\;\; \dot{\mu} = \mu _0\gamma {\Phi _0\over
c^2}(1+3\Omega _\Lambda )H_0. \label{130}
\end{equation}
These two derivatives are negative and very small.  To study the
propagation of the light in a medium whose permittivity and
permeability decrease adiabatically, we must take the Maxwell
equations and deduce the wave equations for the electric field
${\bf E}$ and the magnetic intensity $\bf H$. It is very easy to
show that they are
\begin{equation}
\nabla ^2{\bf E}-{\partial \over \partial t}\left( \mu {\partial
\over \partial t}(\epsilon {\bf E})\right)=0,\;\;\; \nabla ^2{\bf
H}-{\partial \over \partial t}\left( \epsilon {\partial \over
\partial t}(\mu {\bf H})\right)=0, \label{140}
\end{equation}
or, more explicitly,
\begin{eqnarray}
\nabla ^2{\bf E}-{1\over c^2(t)}{\partial ^2{\bf E}\over \partial
t^2} &-&{1\over c^2(t)}
\left({\dot{\mu}\over \mu _0}+{2\dot{\epsilon}\over \epsilon _0}\right)
{\partial {\bf E}\over \partial t} -{1\over c^2(t)}{\dot{\epsilon}
\dot{\mu}\over \epsilon _0\mu _0}{\bf E}=0,\label{150}\\
\nabla ^2{\bf H}-{1\over c^2(t)}{\partial ^2{\bf H}\over \partial
t^2} &-&{1\over c^2(t)} \left({2\dot{\mu}\over \mu
_0}+{\dot{\epsilon}\over \epsilon _0}\right){\partial {\bf H}\over
\partial t} -{1\over c^2(t)}{\dot{\epsilon}\dot{\mu}\over \epsilon
_0\mu _0} {\bf H}=0,\label{160}
\end{eqnarray}
since at present time $\epsilon _{\rm r}=1,\;\mu _{\rm r}=1$.
Because of (\ref{130}), $\dot{\epsilon}/\epsilon _0$ and
$\dot{\mu}/\mu _0$ are of order $H_0=2.3\times 10^{-18}\mbox{
s}^{-1}$, so that the third terms in the LHS of (\ref{150}) and
(\ref{160}) can be neglected for frequencies $\omega \gg H_0$, in
other words for any practical purpose. Note, by the way, that the
last  is a mass term with coefficient of order $10^{-53}\mbox{
m}^{-2}$, {\em i.e.} corresponding to a mass $m_\gamma \simeq
10^{-39}$ MeV or close. This is exceedingly small and can be
neglected also. We are left with two classical wave equations with
time dependent light velocity $c(t)=c+a_\ell t$.
\begin{equation}
\nabla ^2{\bf E}-{1\over c^2(t)}{\partial ^2{\bf E}\over \partial
t^2}=0,\;\; \nabla ^2{\bf H}-{1\over c^2(t)}{\partial ^2{\bf
H}\over \partial t^2}=0. \label{170}
\end{equation}
In order to find the behavior of a monochromatic light beam
according to these two wave equations, we start with the first one
and take ${\bf E}={\bf E}_0\exp[{-i( \kappa z-(\omega
_0+\dot{\omega}t/2)t)}]$, where the frequency is the time
derivative of the phase of {\bf E}, {\em i.e.} $\omega
_0+\dot{\omega}t$. Neglecting the second time derivatives and
working at first order in $\dot{\omega}$ (with $\dot{\omega}t\ll
\omega_0$, $\dot{\omega}\ll \omega _0^2$), substitution in
(\ref{170}) gives $\kappa ^2 =\left[(\omega_0
+\dot{\omega}t)^2-i\dot{\omega}\right]/c^2(t)$. It follows that
$\kappa =k+i \zeta = \pm (\omega_0/
c(t))[1+4\dot{\omega}t/\omega_0](\cos \varphi + i \sin\varphi ),$
with $\varphi =-\dot{\omega}/2\omega_0 ^2,$ so that $k=\pm
(\omega_0 /c)\, (1+\dot{\omega}t/ \omega_0)/(1+a_\ell t/ c)$ what
implies that $k=\pm \omega _0/c$ and the validity of eq.
(\ref{110}), $\dot{\omega} /\omega_0= a_\ell /c$,  as stated
before. Also, $\zeta =-\dot{\omega}/2\omega_0 c =a_\ell /2c^2$.
The wave attenuates in the direction of propagation as $e^{-\zeta
z}$, but since $a_\ell$ is of order $H_0c$, $\zeta ^{-1}$ is of
order of 6,000 Mpc, so that the attenuation of the waves can be neglected. It
is easy to show that to take $k +\dot{k}t$ for the wave vector
leads to $\dot{k}=0$. These results are valid both for the
solutions of (\ref{150}) and (\ref{160}).

This shows that the electromagnetic waves verify eq. (\ref{120}),
so that $k$, and  the wavelength $\lambda$ therefore, remain
constant while the frequency increases with the same relative rate
as the light velocity. Note an important point: in a measurement
of the frequency (necessarily very precise), a blue shift is found
(unrelated to the velocity of the emitter), but observations of
the wavelength fail to find any effect. In other words, the
observation using radio waves can discover the blue shift but the
standard optical observations would see nothing.

The model would give thus a solution to the Pioneer riddle.
However, it seems strange and surprising that this acceleration
would not have been detected thus far. So, in order to proceed, it
is necessary to show that an acceleration of about $10^{-9}\mbox{
m/s}^2$ is small enough to have remained undetected. In other
words, one must consider the experimental tests of spacial
relativity and the equivalence principle as well as other
experiments in which the light speed plays a role. But before
going to that, let us consider in next section some ideas of
general relativity, as they were proposed by Einstein.

\section{Einstein and the speed of light}

Just after proposing his special relativity in 1905, Einstein
realized that his theory was not an end but just a beginning.
Indeed he published later but a few papers on that theory,
starting instead an effort to transcend it that would lead to the
future general relativity of 1916 \cite{Pai84}. Due to their
enormous importance, the interest of the physicists concentrated
in the two relativity theories in their finished form, his work
during the transition period from 1906 to 1915 being much less
known. This explains why, although studied by the historians of
science, its influence on the mainstream of physics had been
scarce. It is however highly interesting, in particular because of
Einstein's discussions on the variation of the light speed in a
gravitational field.

In the last section of a review paper in 1907 \cite{Ein07},
Einstein introduces his principle of equivalence and concludes
that the velocity of light must depend on the gravitational
potential. According to Pais, ``the study of Maxwell equations in
accelerated frames had taught him that the velocity of light is no
longer a universal constant in the presence of gravitational
fields" \cite{Pai84}. It seems that he was not fully satisfied
with his 1907 work so that in 1911 he takes again the question in
a paper entitled ``On the influence of gravitation on the
propagation of light" (after what Pais called ``three and a half
silent years"). In order to apply the principle of equivalence, he
uses two reference frames, K in a gravitational field with
acceleration of gravity $g$, parallel to the $z$-axis in the
negative direction, and K$^\prime$, situated in a space free of
gravitation and moving with acceleration $g$ in the positive
direction of the $z$-axis. In the third and last section ``Time
and velocity of light in the gravitational field", he analyzes the
well known formula for the gravitational redshift $(\nu
_1-\nu_2)/\nu_2=-\Phi/c^2$, where $\Phi= \Phi _1-\Phi _2$ is the
change of the gravitational potential between the points 1 and 2.
He says that this seems to assert an absurdity since it states
that the number of periods per second arriving in 1 is different
that the number emitted in 2.

How  could this be? To answer, he states: ``we cannot regard
$\nu_1$ or $\nu_2$ simply as frequencies since we have not yet
determined the time in system K because [even if we have
synchronized the clocks], nothing compels us to assume that the
clocks U in [points with] different gravitational potentials must
be regarded as going at the same rate" (U is a set of clocks
synchronized as in special relativity). In a somewhat obscure
passage, Einstein argues then that the effect of the gravitational
field is that ``if we measure the time in 1 with the clock U, then
we must measure the time in 2 with a clock which goes $1+\Phi
/c^2$ times more slowly than the clock U when compared with U at
one and the same place". He concludes that ``if we call the
velocity of light at the origin of coordinates $c_0$, where we
take $\Phi =0$, then the velocity of light at a place with
gravitational potential $\Phi$ will be given as
\begin{equation}
c=c_0\left(1+{\Phi \over c^2}\right)." \label{180} \end{equation}
This equation, which was already in the 1907 paper, is number (3)
in reference \cite{Ein11}. Einstein says that it is a first order
approximation. It can be written a bit more
explicitly as
\begin{equation}
c(\Phi ) =c_0\left(1+{\Phi -\Phi _{\rm R}\over c_0^2}\right),
\label{190}
\end{equation}
where $c_0=c(t_0)$ is, and will be from now the speed of light at
$R$, {\it i.e.} the constant that appears in the tables and $\Phi
_{\rm R}$ is a reference potential (at a terrestrial laboratory,
for instance, where the light speed is measured with high
precision).
 Equations (\ref{180})-(\ref{190}) state that $c$ must vary in space, so that
 the deeper (more negative) is the potential, the lower is
the light speed and conversely (at the surface of the Sun, $c$
must be about 2 ppm lower than here at Earth). Note that they are
contrary to the frequent assumptions that $c$ is a universal
constant of nature and that relativity precludes absolutely any
variation of the light speed (see also \cite{Ein11}). It must be
emphasized, furthermore, that eq. (\ref{190}) coincides whith eq.
(\ref{50}) if $\chi =1$ and that we are assuming that $\xi$ is
small.

In two papers in 1912 \cite{Ein12a}, he proceeds with his ideas.
He begins by stating that the light speed is not constant in the
presence of gravitational fields, a result that `` excludes the
general applicability of Lorentz transformations". Turning once
more the nut, he considers the light speed as a field in spacetime
$c({\bf r},t)$ which, in a static situation, verifies $\Delta c=
kc\rho$, $k$ being a constant (it was precisely in this paper and
as a consequence of his thinking on this equation, where he
realized that $\rho$ must include `` the density of ponderable
matter plus the energy density [evaluated locally]". In the second
paper, he reaches the conclusion that the variation of the light
speed affects Maxwell's equations, since $c$ appears in the
Lorentz transformations, what establishes a coupling between the
electromagnetic and the gravitational fields which is not easy to
analyze since the first is generally time dependent. In the second
paper, he proposes, as a dynamical equation for the field $c=c({\bf r}
,t)$, the equation $\Delta c= k[c\rho +(\nabla c)^2/2kc]$
\cite{Pai84,Ron85}.

In 1912, in a reply to a critical paper on relativity by M.
Abraham, Einstein states clearly ``the constancy of the velocity
of light can be maintained only insofar as one restricts oneself
to spatio-temporal regions with constant gravitational potential.
This is where, in my opinion, the limit of the principle of the
constancy of the velocity of light --- thought not of the
principle of relativity
--- and therewith the limit of the validity of our current theory
of relativity, lies" \cite{Ein12b}. Note that what Einstein says
here is that the principle of relativity is not the same thing as
the principle of constancy of light speed: the latter must not be
taken as a necessary consequence of the former. Furthermore, he
admits that the light velocity can depend on $\Phi$, as was the
case with his eq. (\ref{190}).

A final comment to close this section. Because the universe is expanding
the potential $\Phi$ produced by all the matter and energy must increase
so that Einstein eq. (\ref{190}) implies that the speed of light must increase.

\section{The potential of all the universe and the acceleration of
light} Einstein's variable light speed approach to General
Relativity fell into oblivion, superseded as it was by general
relativity that gives a deeper insight on Gravitation.  However
his arguments leading to eq. (\ref{180}) are still right and can
be used in the study of non local effects in weak gravitational
fields, when going beyond special relativity which is only valid
locally  in spacetime. Indeed, special relativity is an
approximation to general relativity which holds good only locally,
more precisely in all the Local Inertial Frames (LIF), {\it i. e.}
free-falling frames in each tangent space. However there is no
reason whatsoever to require that the light speed will be the same
for all the LIF's: on the contrary it can be expected that it will
depend on the particular LIF $\cal L$, so that $c=c({\cal L})$.
For the sake of the argument, let us take a team of physicists in
a spaceship during a travel through regions at which the  potential
is weak, through the Solar System
 or the Galaxy for instance. It they measure locally the light speed they
 could find that it varies adiabatically along the trip. At any moment,
special relativity would  still be valid in a small region around
any spacetime point as a very good approximation. The local value of $c$,
 however, could be different from one region to the other.

This can understood by considering the element of interval in weak
gravity \begin{equation} {\rm d} s^2= e^{2\Phi /c^2}\,c^2\,{\rm d} t^2-{\rm d}
\ell^2\approx (1+2\Phi /c^2)\,c^2\,{\rm d}t^2-{\rm d}\ell
^2.\label{200}\end{equation} Note that $\Phi$ is here the potential of near bodies,
those that produce a non vanishing acceleration $g({\bf r},t)$ at
the observation point. Along a null geodesic one has thus for the
light speed at a generic point $P$, \begin{equation} c(P)=c(R)[1+(\Phi
(P)-\Phi (R))/c^2],\label{210}\end{equation}  $R$ being here a reference
spacetime point in the geodesic. Assume for a moment that the
light speed increases adiabatically as $c(t)=\tilde{c} f(t)$,
where
 $\tilde{c} =c(\tilde{t})$ is the light speed at
some fixed time $\tilde{t}$ in the past, $f=1+\eta(t)$ (so that
$\eta (\tilde{t})=0)$, $\eta (t)$ being  a small and monotonously increasing function of
time. Defining the function $\Pi (t)=\tilde{c}^{\,2}\log
f=\tilde{c}^{\,2}\log [c(t)/\tilde{c}]\approx \tilde{c}^2\eta (t)$, the interval (\ref{200})
can be written as \begin{equation} {\rm d} s^2= e^{2(\Phi +\Pi )
/\tilde{c}^{\,2}}\,\tilde{c}^{\,2}\,{\rm d} t^2-{\rm d} \ell^2,
\label{220}\end{equation} at first order. It is clear
then that $\Pi$, which is space independent but time
dependent, can be interpreted as a uniform gravitational potential
of global character. Since (\ref{200}) and
(\ref{220}) give equivalent descriptions, it turns out that an
adiabatic acceleration of light is equivalent to the gravitational
potential of all the universe in the following sense. At any time,
we can use either (i) eq. (\ref{200}) with a time dependent  light
speed $c(t)$ and no potential $\Pi$, or (ii) eq. (\ref{220}) with
a constant value of the light speed $\tilde{c}$ and with the
potential $\Pi (t)$. This shows that $\Pi (t)$ {\it is the change
of the gravitational potential of all the universe since the time $\tilde{t}$}.
If in the second option we take
$\tilde{t}=t_0$ and, consequently $\tilde{c}=c_0$ ({\it i. e.} the
value in the tables), then $\Pi (t_0)=0$ but $\dot{\Pi} (t_0)
>0$ because the
galaxies are separating in the universal expansion, this implying
that light must be accelerating.

\subsection{Is time accelerating?}As shown in \cite{Ran04}, there
is a third (and intriguing) interpretation. In section 2, the quantity
$a_{\rm t}$ such that $a_\ell =a_{\rm t}c$,
was introduced and calculated to
be $a_{\rm t}=2.2\times 10^{-18}\simeq H_0$. Anderson {\it et al} mentioned in \cite{And98} that
the Pioneer acceleration could be related to a ``clock acceleration
$a_{\rm t}$ of $2.8\times 10^{-18}\mbox{ s/s}^2$",
such that ``$a_{\rm P}=a_{\rm t}c$"
(compare with eq (\ref{110}) of this work), that
``would appear as a nonuniformity
of time; {\it i.e.} all clocks would be changing with constant acceleration".
However, although the authors say ``We have not yet been able to rule out this possibility.",
they do not further elaborate this idea, at least there, perhaps because they say that some results
``rule out the universality of the time-acceleration model".

It is encouraging  that the calculation in section 2 of
this paper gives for $a_{\rm t}$ a value close to the Hubble parameter and to the
one deduced in \cite{And98} from the value of the blue shift which was interpreted as the acceleration $a_{\rm P}$.

Note that this can be formulated by defining an accelerated time
$t_{\rm cosm} $ (measured by atomic clocks) so that
$t_{\rm cosm}=t+H_0t^2/2$ (since $a_{\rm t}=H_0$) \cite{Ran04}. As is easy to see,
although the light speed would
increase with respect to time $t$, it would be constant with
respect time $t_{\rm cosm}$. What could this mean? A possibility
would be that $t$ is a  parametric time, coinciding with the usual
time in Newtonian and even in special and general relativity,
while $t_{\rm cosm}$ could be the cosmological time (note that
it would reverse its arrow in a contraction of the universe in which
the time derivative $\dot{\Phi} _{\rm av}/c^2=H$ would be negative.) The difference
between these two kinds of time is important and complex, a
theory embodying the dynamics of time would be then necessary \cite{Tie02,Bar98}.
The Pioneers anomaly would show this duality of times and its
understanding would need a good theory for the dynamics of time.
All this will
be considered in a forthcoming paper.

Another important point must be emphasized. As shown in
\cite{Ran03a,Ran03b,Ran03c} the potential $\Pi$ due to all the
universe is much bigger that the potentials due to the local
inhomogeneities, as the Galaxy for instance, so that one could
fear that the a linearization or a Newtonian approximation would not
valid. But we see here that, a large value of $\Pi$
can be eliminated just by changing accordingly the light speed. We
can do away with the potential $\Pi$ in the expression of the
interval, by the simple procedure of using the current value of
$c$.  The linear approximation is thus thus quite acceptable.
However as $\Pi =\Pi (t)$, its time derivative, although very
small, does not vanish and, consequently, the value of $c$ must
vary.

\section{First summary: spacetime variations of \mbox{\boldmath$e$},
\mbox{\boldmath$c$}, \mbox{\boldmath$\alpha$} and \mbox{\boldmath$e$}}

In the following, the predictions of this model will be compared
with the experimental tests of special relativity and the
equivalence principle (see the classic books by Dicke
\cite{Dic68}, Misner, Thorne and  Wheeler \cite{Mis73} and Will \cite{Wil93}). Taking into account the
previous considerations,  it follows from eqs. (\ref{40})-(\ref{60})
that the difference between the values of $e$, $c$, $\alpha$ and
$m$ at two points are, at first order, \begin{equation} {\Delta e\over
e}={\chi + \xi\over 2} \, {\Delta \Phi \over c^2},\quad {\Delta
c\over c}=\chi \, {\Delta \Phi \over c^2},\quad {\Delta \alpha
\over \alpha }=\xi \, {\Delta \Phi \over c^2},\quad {\Delta m\over
m}=(-\chi +\xi)\, {\Delta \Phi \over c^2}, \label{230} \end{equation} the
last equality coming from the assumption that the electron mass
verifies $mc^2\propto e^2$ \cite{Sch66}. Note that the Planck
constant $\hbar$ is assumed to be invariant, not depending
therefore on the potential. Remember that we assume that
$\chi =1$, $\xi \lesssim 10^{-5}$.

In the following the predictions based in this model, more
precisely in equations (\ref{230}) will be compared with the
experimental tests of the equivalence principle and special relativity.

\section{Experiments on the gravitational redshift}
General Relativity predicts a shift, {\it the gravitational
redshift}, of a light ray as it travels from an emitter 2  to a
receiver 1. It is emitted with frequency $\nu _2$ and received with frequency $\nu_1$.
The shift is given, at first order in the potential, as
\begin{equation}
\nu _{\rm 1} =\nu _{\rm 2}\left[1- {\Phi _{\rm 1}-\Phi _{\rm
2}\over c^2}\right],\label{240}\end{equation} or, equivalently \begin{equation}
 {\nu_{\rm 1}-\nu_{\rm 2} \over \nu_{\rm 2}} = -{\Phi _{\rm 1}-\Phi _{\rm 2}\over
 c^2} \label{250}
\end{equation}
(if the light goes upwards at Earth surface (or it goes away from
the Sun) its frequency decreases and conversely). Note that (\ref{240})
is a consequence of the weak equivalence principle.
When looking for
eventual violations of eq. (\ref{250}) it is customary to write it
as \begin{equation} {\Delta \nu \over \nu} =-(1+a){\Delta \Phi \over c^2}.
\label{260} \end{equation}
The prediction of general relativity is $a=0$,
so that a nonvanishing $a$ would indicate a failure of the
equivalence principle or of the presence of another effect. The
best bounds for $|a|$ are  $|a|\leq 10^{-2}$ for the nuclear gamma ray
spectrum by Pound and coworkers \cite{Pou60,Pou65}  and
$|a|\leq 2\times 10^{-4}$ obtained by Vessot, Levine
and coworkers in the hyperfine spectrum of Hydrogen
 \cite{Ves79,Ves80}.

 In the usual interpretation of this experiment, $\nu _1$ is the
 frequency of the radiation of a particular line that
 reaches the laboratory and $\nu _2$ is the frequency of the same line when emitted at the source 2,
 which is assumed to be equal to the frequency of the same radiation when emitted at the laboratory 1.
It is accepted that  the photons lose energy when
going to higher potentials and conversely, just as a particle that
loses kinetic energy when gaining potential energy. On the other hand, what happens
in this model is that any particular line is emitted
with lower frequency if the potential is
deeper (and conversely), travelling afterwards with constant frequency
from the emitted to the receiver. Indeed, since the frequency and
wavelength of an emitted line are functions of the electron charge
$e$ and mass $m$ and the light speed $c$, that frequency and that
wavelength depend on space and time (because these quantities are
slightly variable across the spacetime, see eqs.
(\ref{40})-(\ref{60}) and note that $mc^2\propto e^2$). As a
consequence, in this Newtonian model, the frequencies of the
spectra are indeed space and time dependant.
 In the case of the redshift experiments, the time variation of the potential can be neglected.
 The space variation is then $\Delta \Phi = \Phi _2-\Phi _1$.

Let us find now which is the frequency and wavelength predicted by
this model in the main experiments performed until now to test the
gravitational redshift. We will consider four possibilities depending on
whether the lines are produced in (i) ordinary atomic spectrum,
(ii) fine structure atomic transitions, (iii) gamma nuclear
transitions and (iv) hyperfine transitions. In each case, the
production of a beam of light at space-point $S$ and its reception
at point $R$ will be considered.

It is convenient to use the following notation. The indices $R$ and $S$, the initials of ``reference"
and ``source", refer to the laboratory at Earth and to
the emitter. $\nu _S$ and $\lambda _S$ (resp. $\nu _R$ and $\lambda _R$) will denote
the frequency and wavelength of the radiation as is emitted at $S$ (resp. at $R$);
 $\nu _R^{\,\prime}$ and $\lambda _R^\prime $ are the same
quantities of the radiation emitted at $S$ when it arrives to $R$;
$c_S$ and $c_R$ are the value of the light speed at the source and at the laboratory.

With this notation, eqs. (\ref{240})-(\ref{250}) are written
$$\nu ^\prime _R = \nu _R\left[1+ (1+a){\Phi _S-\Phi _R\over c^2}\right]\qquad
{\nu ^\prime _R- \nu _R\over \nu _R}= (1+a){\Phi _S-\Phi _R\over c^2}$$

Remember that we are using $\chi =1$ and $\xi =
O(10^{-5})$ or smaller and that no variation of $\hbar$ is considered here.

\subsection{Ordinary spectrum} The frequencies of this
type of spectrum verify the proportionality
$$ \nu \propto mc^2h^{-1}\alpha ^2$$
For instance, they are equal to $\nu
_{nm}=mc^2\alpha ^2h^{-1} (1/2n^2-1/2m^2)$ in the Hydrogen spectrum.

Taking into account eqs. (\ref{230}), it turns out that the
frequency of a particular line emitted at a spacetime point $S$
verifies $\nu _R\propto me^4$, $m$ and $e$ being the values of the
electron charge and mass at $S$. Therefore, under a change of the
potential, it suffers the relative change $\Delta \nu /\nu = (\chi
+3\xi )\Delta \Phi /c^2$ so that one has \begin{equation} \nu _S=\nu
_R[1+(\chi +3\xi)\Delta \Phi /c^2],\qquad \nu _R^\prime =\nu _S\label{280}\end{equation} with $\Delta
\Phi =\Phi (S)-\Phi (R)$. Note that the frequency of the radiation
does not change during its flight from $S$ to $R$ if the effect of
the time variation of $\Phi$ can be neglected, as it happens in
the redshift experiments performed up to now (remember that the
space is here an optical medium with refractive index $n=n(\Phi)$
given by eq. (\ref{70})). The frequency of the radiation received
at $R$ is thus $\nu _R^{\, \prime}=\nu _S$. Note also that $\nu
_R^{\,\prime} <\nu _R$ (resp. $\nu _R^{\,\prime}
>\nu _R$ if the wave travels in the direction of increasing (resp.
decreasing) potential, so that there is a red shift (resp. a blue
shift). Concerning the wavelength and since $c_S =c_R [1+\chi
\Delta \Phi /c^2]$, one has for the wavelength of the radiation
emitted at $S$ and received at $R$, respectively \begin{equation} \lambda _S
={c_S\over \nu _S} =\lambda _R [1-3\xi\Delta \Phi /c^2],\quad
\lambda _R\prime = {c_R\over \nu _R^{\,\prime}}=\lambda _R[1-(\chi
+3\xi )\Delta \Phi /c^2].\label{290}\end{equation} The consequence is that
this model predicts that the variation is given by eq. (\ref{260})
with $|a|=3|\xi| \leq 4\times 10^{-5}$. The best bound in ordinary
spectrum, obtained by Braoult by observing solar spectrum lines is $|a|\leq 5\times 10^{-2}$. So the prediction of this Newtonian model
is not in conflict with the observations.

In conclusion the prediction of the model is
exactly the same as predicted by the weak equivalence principle
and by General Relativity if $\xi=0$. It is equal within the experimental
margin if $|\xi |\leq 10^{-5}$.

\subsection{Fine structure spectrum}
The fine structure must be considered also. The difference between
the frequencies within the same multiplet verifies $\nu \propto
mc^2\alpha ^4\propto me^4\alpha ^2$. It is very easy to see,
following the same analysis as in the preceding subsection, the
only difference being that instead of (\ref{280}) and (\ref{290})
one has \begin{equation} (\nu _R^{\,\prime}=)\nu _S=\nu _R[1+(\chi
+5\xi)\Delta \Phi /c^2],\quad \lambda _R\prime = {c_R\over \nu
_R^{\,\prime}}=\lambda _R[1-(\chi +5\xi )\Delta \Phi
/c^2].\label{300}\end{equation} which corresponds to $|a|\leq 8\times
10^{-5}$.

\subsection{Nuclear gamma ray spectrum}
A very important, well known and precise experiment was performed
by Pound, Rebka and Snyder \cite{Pou60,Pou65}, by using the
Mössbauer effect to measure the redshift of a line of 14.4 keV
gamma rays of Fe$^{57}$. The source and the receiver were at the
the bottom and the top of a tower, respectively, in the Jefferson
Laboratory of Harvard University, at the distance of 22.5 m.  To
measure the shift, They observed the absorption of photons in a
target containing Fe$^{57}$, by setting the source in motion to
compensate the gravitational redshift with a Doppler effect and
measuring the velocity at which the photons were absorbed. Their
result was that $a\leq 10^{-2}$. From the semiempirical mass formula
of von Weizs\"acker the energy of the transition is proportional to $e^2$.
  Following then the same arguments as
before and using (\ref{230}), it is easy to show that, instead of (\ref{280}) and
(\ref{290}) one has \begin{equation} (\nu _R^{\,\prime}=)\nu _S=\nu _R[1+(\chi
+\xi)\Delta \Phi /c^2],\quad \lambda _R\prime = {c_R\over \nu
_R^{\,\prime}}=\lambda _R[1-(\chi +\xi )\Delta \Phi
/c^2].\label{310}\end{equation} which corresponds to $|a|\leq 2\times
10^{-5}$, well below the bound $|a| \leq 10^{-2}$ implied by this
experiment.

Note that, according to the interpretation of this model, the
photons were emitted with a lower frequency at the bottom that
they would have been at the top, this frequency being enhanced by
the motion of the source to produce the resonant absorption. We
see that the model agrees with the Pound-Rebka-Snider experiment.

\subsection{Hyperfine spectrum}
The best experimental bound obtained up to now for the
gravitational redshift was obtained by Vessot and Levine
\cite{Ves79,Ves80} measuring the frequency variation of the 1,420
MHz line of the hyperfine spectrum of Hydrogen between a spaceship
in vertical motion up to a height of 10,000 km and a laboratory at
Earth surface. Their result is $$|a| \leq 2\times 10^4.$$ The
energy levels of that hyperfine structure have the expressions
$$E= {\mu _0\over 4\pi} g_p \mu _N\mu _B K\left({L(L+1)\over J(J+1)}
\langle {1\over r^3}\rangle +{2\over 3} R^2(0)\right),$$ where
$K=F(F+1)-I(I+1)-J(J+1)$, $\mu _N=e\hbar /2M$ and $\mu _B=e\hbar
/2m$ are the nuclear and the Bohr magnetons, $g_p=2.79$, $J,I,F$
are the electron, nucleus and total angular momentum and the
average value of $1/r^3$ and the  value of the radial function at
the origin appear in the last factor \cite{Sak67}. From
(\ref{230}) we have $\Delta \mu _B = (3\chi -\xi) \Delta \Phi
/2c^2$. Let us assume that $\Delta g_p/g_p=(\chi +\xi)\Delta \Phi
/c^2$, what means that $g_p$ varies approximately as $e^2$, as can
be expected. All this means that the frequencies in the hyperfine spectrum are
proportional to
$${\mu _0\over 4\pi}g_p\mu _N\mu _B\hbar ^2\left({me^2\over 4\pi \epsilon _0\hbar ^2}\right)^3=
{g_p\over 4\pi}\, {m^2\over M\hbar}\,\left({e^2\over 4\pi \epsilon
_0\hbar c}\right)^4c^2\propto g_p{m^2\over M} \alpha ^4c^2,$$
where the relation $\mu _0=1/c^2\epsilon _0$ has been used. It
follows that
$$\nu _S= \nu _R\left(1+[(\chi +\xi) +2(-\chi+\xi )+4\xi +2\chi]
{\Delta \Phi \over c^2}\right)=\nu _R\left(1+(\chi +7\xi) {\Delta
\Phi _{\rm s}\over c^2}\right).$$ Using the same arguments as for the
ordinary spectrum, it is found that in this case $a=7\xi
\simeq 9\times 10^{-5}$, which is close from below to the
Vessot-Levine bound. However, remember that there are arguments to
suggest that $\xi$ is probably smaller, perhaps even zero.

\subsection{Frequency from distant sources}
The time variation of $\Phi$ can not be neglected in the case of the
radiation coming from distant sources, contrary to what happens in the
experiments referred to in sections 7.1-7.4. It could seem at first sight,
therefore, that the increasing in time of the frequencies of the radiation
during its travel to Earth must pose a problem, since the radiation coming from distant
sources would be blueshifted with $\Delta \nu =\nu a_{\rm t} D/c^2$,
where $D$ is the distance from the source, contrary to what is observed.
The previous considerations in this section indicate that there is no such a problem.
Let us consider a line which has frequency $\nu_0$ here when it is emitted here.
It follows from sections 7.1-74 that the frequency with which it is emitted at a source $S$,
 placed at distance $D$, is
$$\nu _{S}=\nu _0\left[1+(\chi +b\xi ) \left({\Phi _{\rm av}(t)
-\Phi _{\rm av}(t_0)\over c^2}\right)\right]=\nu _0\left[1-(\chi +b\xi )a_{\rm t}{D\over c}\right],$$
where $t_0-t=cD$ and $\Delta \Phi /c^2 =-a_{\rm t}D/c$. As shown before, the numerical coefficient $b$ is equal to 3 (ordinary spectrum),
5 (fine structure spectrum), 1 (gamma rays) and 7 (hyperfine spectrum).
It is clear that it is emitted at $S$ with a lower frequency but, in the travel to Earth during the time $D/c$,
it increases in  a relative amount $\chi a_{\rm t}D/c$, so that it will be received at Earth with the value
$$\nu _R^\prime = \nu _0 \left[ 1-b\xi a_{\rm t}{D\over c}\right].$$
The cosmological redshift must be added, what decreases further the frequency
by a relative amount $H_0D/c =a_{\rm t}D/c$, i. e. to $\nu _0[1-(1+b\xi )H_0D/c]$. This means that the effect of the quantum vacuum
used in this model increases the observed cosmological redshift by the factor $(1+b\xi)$.
Since $|b\xi| \lesssim 10^{-5}$ and could be nil, this is quite unobservable.
 This property can be expressed by saying that the changes in the frequency
  of a line due to the time changes in the emission and in the propagation
  of the radiation until Earth cancel each other, except for a very small and unobservable multiplicative factor.

The conclusion of this section is that, if there is no cosmological
variation of the fine structure constant as some papers suggest
(for instance \cite{Sri04}) then $\xi =0$, the predictions of the model
being the same as those of the weak equivalence principle
 and fully compatible therefore with the experiments.
If there is a variation of $\alpha$ with
the value obtained in \cite{Web01} then $\xi \simeq 10^{-5}$,
 the model being still compatible with the observations.
If that variation
is smaller than the  first estimation by Webb et al \cite{Web01}, the
difference between the predictions of the model
and of the weak equivalence principle would be smaller also.

\section{Bending of a light around the
Sun}

Let us consider a light beam that grazes the Sun, its mass being $M$ and
its radius $R$. Because the refraction index of the quantum vacuum
decreases as the distance to the center increases, the trajectory of a
light ray is a curve $y=y(x)$ that that bends towards the center
of the star. The time variation of the light speed can be neglected here.
The velocity of light and the refraction index take
the form $$ c(r) = c\left(1-\eta {R\over r}\right),\quad n(r)
= \left(1+\eta {R\over r}\right), $$ where $\eta =GM/c^2R=2.1
\times 10^{-6}$ and $c(\Phi \!=0)=
c(1-\Phi _{\rm R})$ has been rebaptized as $c$.

The ray trajectories are the solutions to the variational problem
\begin{equation} \delta T[y(x)] =\delta \int _1 ^2{n(r) \over c}(1+y^{\prime
2})^{1/2}dx =0 \label{320} \end{equation} or \begin{equation} \delta T[y(x)] =\delta
\int _1 ^2{(1+y^{\prime 2})^{1/2} \over c}\left(1+\eta {R\over
r}\right)dx =0. \label{331}\end{equation}
 The corresponding Euler-Lagrange equation is
\begin{equation} {d\over dx}\left[{y^\prime \over (1+y^{\prime
2})^{1/2}}\left(1+\eta {R\over r}\right)\right]+ \eta (1+y^{\prime
2})^{1/2}{yR\over r^3}=0. \label{340} \end{equation} We wish to obtain the
solution with initial data $y(0)=R, y^\prime (0)=0$. As $\eta$ is
small, the solutions can be expanded in series $y(x)=y_0(x)+\eta
y_1(x) +\cdots$.

 At  order zero the equation (\ref{340}) is $$
{d\over dx}\left[{y_0^\prime \over (1+y_0^{\prime 2})^{1/2}}\right]=0.
$$ which says that $y_0^\prime $ is constant. The zero order
approximation is thus $y_0(x)=R ,y^\prime (x)=0$ so that at order 1:
$$ {d\over dx} y_1^\prime =-(1+y_0^{\prime 2})^{1/2} {y_0R\over
(x^2+y_0^2)^{3/2}} ={-R^2 \over (x^2+R ^2)^{3/2}}. $$ The solution
to with the prescribed initial data is $$ y_1^\prime (x) = -\int
_0^x{R^2dx\over (x^2+R^2)^{3/2}} = -{x\over (R^2+x^2)^{1/2}}, $$
so that $y_1(x)=-(R^2+x^2)^{1/2}$ as can be seen easily by means
of the change $x=R\tan \beta$. It follows that $y(x)=R-\eta
(R^2+x^2)^{1/2}$. The bending angle is $\phi =2|y^\prime (\infty)|$,
 i.e \begin{equation} \phi =2\eta =2{GM\over c^2R}= 0.875^{\prime \prime}.
\label{3350} \end{equation} The well tested prediction of General Relativity
is twice as much, {\it i. e.} $\phi =1.75^{\prime \prime}$.
This means that the prediction of this model is exactly the same as that of the
weak equivalence principle.

\section{Time delay of radar signals}
According to general relativity, light waves take a longer time to
traverse a distance in the gravitational field of a massive body,
the Sun for instance, than they would in Newtonian physics with
constant $c$. It is said that they suffer a time delay. This
effect, predicted by Shapiro in 1964 \cite{Sha64,Mis73,Wil93}, is
one of the classic tests of general relativity. Let a beam of
radar waves be emitted from Earth and let it be reflected back
from a reflector, somewhere in the Solar System. More precisely,
let us consider the case of a signal grazing the Sun surface. Let
$a_{\rm E}$ (resp. $a_{\rm R}$) be the distance from the emitter
at Earth (resp. the reflector, say a planet) and the point where
the trajectory touches the Sun, to be approximated as the radii of
their orbits, and let $M,\, R$ be the mass and radius
of the Sun. The well known prediction of general relativity is
that the round trip time is \begin{eqnarray} \tau _{\rm GR}&=&|g_{00}|_{\rm
Earth}^{1/2}\,2\int _{-a_{\rm
E}}^{a_{\rm R}}\left[1+{2GM \over c^2\sqrt{x^2+R ^2}}\, {{\rm d} x\over c}\nonumber\right]\\
&=&2{a_{\rm E}+a_{\rm R}\over c}\,\left(1-{GM \over
c^2\sqrt{a_{\rm E}^2+R ^2}}\right) \label{400}\\
&+& {4GM\over c^3}\log \left[{(a_{\rm R} +\sqrt{a_{\rm
R}^2+R^2})(a_{\rm E} +\sqrt{a_{\rm E}^2+R^2})\over
R ^2}\right] \nonumber \end{eqnarray}   In this model, there is also
a delay caused by the smaller speed of light near the Sun. The
time variation of that speed can be neglected and the space
variation takes the form
$$c({\bf r})=c_0\left[1-{GM \over c_0^2}\left({1\over r}-{1\over \sqrt{a^2_{\rm
E}+R^2}}\right)\right].$$ (Note that $x^2+R ^2 =r^2$).
It follows that the time delay due to the variation of the speed
of light is in this model \begin{eqnarray} \Delta \tau _{\rm model}
&=&2\int_{a_{\rm E}}^{a_{\rm R}} \left( c({\bf
r})^{-1}-c_0^{-1}\right) {\rm d} x\nonumber \\
&=&{2GM\over c^3}\log \left[{(a_{\rm R} +\sqrt{a_{\rm
R}^2+R^2})(a_{\rm E} +\sqrt{a_{\rm E}^2+R^2})\over
R ^2}\right], \label{410}\end{eqnarray}  The Shapiro time delay is
$\Delta \tau _{\rm GR}=\tau_{\rm GR}-\tau _{\rm classic}$. Since
$c$ in (\ref{400}) is $c_\infty$, the speed of light at infinite
distance (where $\Phi =0$), if $c_0$ is its value at Earth, then
$c=c_0(1-GM /c_0^2\sqrt{a_{\rm E}^2+R^2})$. As $\tau
_{\rm classic} = 2(a_{\rm E}+a_{\rm R})/c_0$, it is equal to the
second line in (\ref{400}), at first order in the variation of
$c$, so that the time delay in this non relativistic model is one
half the prediction of general relativity. This model predicts
again the same result as the weak equivalence principle.

\section{Other tests of the weak equivalence principle}
\subsection{Effect on the electromagnetic mass of bodies}

 In this model, the electromagnetic mass of a body varies
  in a way that depends on its chemical composition,  what
would lead to a violation of the weak equivalence principle (see
\cite{Uza02,Bek02}.  However, the effect would be too small to be
a matter of concern, as we will see now. Moreover it might be worth
to recall the much quoted sentence by  J.L. Synge in his classic
treaty ``Relativity: the General Theory" \cite{Syn60}: ``The
principle of Equivalence performed the essential office of midwife
at the birth of General Relativity, but, as Einstein remarked, the
infant would never have got beyond its long-clothes had it not
been for Minkowski's concept. I suggest that the midwife be now
buried with appropriate honors."

 The mass of a body includes a part of
 electromagnetic origin. In fact, the von Weizs\"acker semiempirical
mass formula  tells us that there is a Coulomb contribution to the
rest energy of a nucleus given by the expression
\begin{equation}
m_ {\rm C}\, c^2= a_{\rm C}Z(Z-1) A^{-1/3}, \mbox{ with }\;\; a_{\rm C}=
{3\over 20\pi \epsilon _0}\,{e^2\over r_0}\simeq 0.6 \mbox{
MeV}/c^2\,\;\; (r_0\simeq 1.5\mbox{  fm}), \label{420}
\end{equation}
 plus the electromagnetic mass of each of the protons. If the mass
of a nucleus is written as  $m= m_0+m_{\rm C}$, then  $u=m_{\rm
C}/m$ is of the order $5\times 10^{-3}$ for the heavier nuclei and
about
 $10^{-3}$ or smaller for the lighter.
 It follows then form (\ref{320}) that the difference of the value of the mass between
 two points due to the variation of $e$ and $c$ is in the model
\begin{equation}
\Delta m =\Delta m_{\rm C}= (-\chi +\xi ) m_{\rm C}\Delta \Phi /c^2
=(-\chi +\xi ) um\Delta \Phi /c^2. \label{430}
\end{equation}
 This implies that the electromagnetic part of the mass of a nucleus
changes from one space point to another in a way that depends on
$\Phi$, on the mass number $A$ and on the atomic number $Z$. In
the case of a macroscopic body, the variation of the mass $\Delta
m$ depends on its chemical composition and on the direction of the
displacement between the two points. This implies that its mass
is, in this model, a function of space and time of the form
$m+\Delta m(x,y,z,t)$. Although $\Delta m$ is a scalar, not a
tensor, there is an anisotropy in the sense that its derivative
does depend on the direction.  It must be emphasized, however,
that $\Delta m$ would be so small that it could not be appreciated
at Earth, as shown below.

During its motion the mass change in time, so that its derivative
is \begin{equation} {{\rm d} m\over {\rm d} t} =(-\chi +\xi ) m_0u {\nabla \Phi \cdot {\bf
v}+\partial _t\Phi \over c^2}. \label{440} \end{equation} Newton's 2nd law
$\dot{\bf p}={\bf F}$ takes the form \begin{equation} {{\rm d} (m{\bf v})\over
{\rm d} t}=m{\bf g},\;\;\mbox{ in other words: } m{{\rm d} {\bf v}\over
{\rm d} t} =m{\bf g}- {{\rm d} m\over {\rm d} t}{\bf v} \label{450} \end{equation} It
turns out that there is an extra force $-\dot{m}{\bf v}=
-m_0u(-\chi +\xi ) \left(\nabla \Phi \cdot {\bf v}+\partial _t\Phi
\right) {\bf v}/c^2$. However, it is extremely small at
terrestrial scale, quite unobservable as will be seen.

The term in $\partial _t\Phi/c^2$ can be neglected in laboratory
experiments. In fact, as shown before, it is equal to $a_{\rm t}
\simeq 2.3 \times10^{-18}\mbox{ s}^{-1}$. Let us see what happen with the
gradient term. The value of $\Phi$ at a terrestrial laboratory is
the addition of the contributions of the Earth, the Sun, the Milky
Way and the rest of the universe $\Phi =\Phi _E+\Phi _S+\Phi
_{MW}+\Phi _U$, with self-explaining notation. The last  one is
the larger but has the smallest gradient. We concentrate therefore
on the other three. The three accelerations are directed towards
the centers of the Earth, the Sun and the Milky Way, respectively,
so that the gradients of the three potentials have moduli $|\nabla
\Phi|=GM/R^2$, $M$ and $R$ being   the corresponding masses and
distances.  In a terrestrial laboratory, it turns out that
(assuming that $M_{MW}=10^{11}M_S$) \begin{equation} {|\nabla \Phi _E|\over
c^2}\simeq 10^{-16}\mbox{ m}^{-1},\;\;\;{|\nabla \Phi _S|\over
c^2}\simeq 10^{-19}\mbox{ m}^{-1},\;\;\;{|\nabla \Phi _{MW}|\over
c^2}\simeq 10^{-27}\mbox{ m}^{-1}, \label{460} \end{equation}
 If a body is displaced the vector $\bf h$, the relative variation
 of its mass is therefore the sum of the corresponding three terms,
which verify \begin{eqnarray} \left.{|\Delta m|\over m}\right|_E &\leq& (\chi -\xi )
h\times 10^{-19},\;\;\; \left.{|\Delta m|\over m}\right|_S \leq
(\chi -\xi ) h\times 10^{-22},
\nonumber\\ \left.{|\Delta m|\over m}\right|_{MW}
\!\!\!\!\!&\leq &(\chi -\xi ) h\times 10^{-30}, \label{470} \end{eqnarray} where $h$ is the
number of meters (remember that $\chi =1$, $\xi \leq 10^{-5}$). The largest change of the mass is the direction
to the center of the Earth. Note that this contribution from Earth
vanishes if $\bf h$ is horizontal. In any case the variation of
$m$ along a trajectory can be neglected at Earth.

\subsection{Free fall} First and in order to understand the problem,
let us consider the free fall along the vertical  of a small body
at the Earth surface, with initial data $v=\dot{z}=0$, $z=z_0$. As
explained before \begin{equation} m=m_0 \left\{1+u\left[1+ (-\chi +\xi ){\Phi (z)-\Phi
(z_0)\over c^2}\right]\right\}, \label{480} \end{equation} where  $m=m_0+m_C$ at the
point $z_0$.
 It turns out therefore that the extra force is very small. In fact its ratio to $mg$ is
\begin{equation} {\dot{m}v \over mg}\simeq u(-\chi +\xi ) \times
10^{-15}\sqrt{z_0-z},\;\;\; (\simeq 10^{-17}\;\; \mbox{ if }z-z_0=
100\mbox{ m}). \label{490} \end{equation}

This force is extremely small at the terrestrial scale, in fact it
is much smaller than the air resistance $F_{\rm S}$. To be specific,
taking the case of a sphere of aluminium with $R=10$ cm and the
Stokes expression $F_{\rm S}=6\pi R\eta v$, it turns out that
$\dot{m}v/F_S\simeq \times 10^{-11}\sqrt{z_0-z}$. The effect is
certainly unobservable.

\subsection{Experiments of E\"otv\"os type.}
After some work by Bessel in the 19th century (and by Newton
before!), E\"otv\"os  and coworkers performed a series of  famous
experiments with a torsion balance to see if the weak equivalence
principle (WEP) is valid (the best one in 1922). To analyze its
experiment let us write the passive gravitational mass $m_P$ as
$m_{\rm P}=m_0+ m_{\rm C} +\Delta m_{\rm C}$, assuming that $m_I=m_0+ m_{\rm C}$ is the
inertial mass. The term $\Delta m_{\rm C}=\Delta m=(-\chi +\xi ) um\Delta \Phi
/c^2$ is the one that breaks the Equivalence Principle, since it
changes in space in a way that depends on the composition of the
body. The time variation of $\Phi$ can be neglected again here.
It follows that \begin{equation} m_{\rm P} = m_0 + m_{\rm C} - (\chi -\xi )
m_0u{\Phi
({\bf r})-\Phi _{\rm R}\over c^2}. \label{500} \end{equation} Note that
the equality $m_{\rm P}=m_0+m_{\rm C}$ holds exactly at R and that
we are working at first order in $u$. The accelerations of the two
bodies are \begin{equation} a_1 =\left[1-u_1(\chi -\xi ) {\Phi ({\bf r}_1)-\Phi _{\rm R}
\over c^2}\right] g,\;\;\; a_2 =\left[1-u_2(\chi -\xi ) {\Phi
({\bf r}_2)-\Phi _{\rm R})\over c^2}\right] g, \label{510} \end{equation}
where ${\bf r}_1, {\bf r}_2$ are the positions of the two arms of
the torsion balance and $g$ the acceleration of gravity. To test
WEP one has to measure the E\"otv\"os ratio \begin{equation} \eta \equiv
2\left|{a_1-a_2\over a_1+a_2}\right|. \label{520} \end{equation} From
(\ref{510}), its value is in this case \begin{equation} \eta =(-\chi +\xi )
\left[u_1{\Phi ({\bf r}_1)-\Phi _{\rm R}\over c^2}- u_2{\Phi
({\bf r}_2)-\Phi _{\rm R}\over c^2}\right] \label{530} \end{equation} Let
$u_1>u_2$. Taking $\Phi ({\bf r}_2)=\Phi _{\rm R}$, we have \begin{equation} \eta =(-\chi +\xi )
u_1{\nabla \Phi \cdot {\bf h}\over c^2}, \label{540} \end{equation} where
${\bf h}={\bf r} _1-{\bf r} _2$. In an E\"otv\"os type experiment $h< 1$ m.
It turns out that \begin{equation} |\eta |\leq h \times 10^{-18} \label{550}
\end{equation} where $h$ is in meters.
  The best experiments by Roll, Krotkov and Dicke \cite{Rol64} and by Braginsky
 and Panov \cite{Bra72} establish as an upper bound for this quantity $10^{-11}$
and $10^{-12}$. The effect used in this model can not be detected
in an E\"otv\"os experiment.

\section{Experimental tests of the local Lorentz invariance}
\subsection{The Hughes-Drever experiments}
The experiments by Hughes {\it et al} \cite{Hug60} and Drever
\cite{Dre61}, following a suggestion by Cocconi and Salpeter
\cite{Coc58}, were devised as tests of the Mach principle. They
establish bounds to the part of mass of a body $\Delta m$ which is
anisotropic. According to that principle and since the matter in
our galaxy is not distributed isotropically around the Earth, the
mass of a body might depend here on the direction of its
acceleration, what would imply in the strict sense  that the
inertial mass would be a tensor $M_{ij}$ with three different
eigenvalues $m+\Delta m_i$. Let $\Delta m$ be the greatest
$|\Delta m_i|$. By studying the Zeeman effect in excited states of
atoms and nuclei when the magnetic field varies with respect to
the direction to the center of the Milky Way as the Earth rotates,
Hughes and Drever established that
\begin{equation}
{\Delta m\over m} < 10^{-20},\;\;\mbox{ and }\;\; {\Delta m\over
m} < 5\times 10^{-23},\mbox{ respectively.} \label{560}
\end{equation}
 It must be emphasized, however, that the mass of a body does not
depend in this model on the direction of the acceleration. In this
sense, there is no anisotropy in it. However, this model can be
considered as a realization of the Mach principle (since some properties of bodies depend in it
of their interaction with all the universe) and
embodies  another kind of anisotropy to which the Hughes-Drever
bounds can be applied.
 As a measure of such anisotropy we must take the value of the changes
of $\Delta m$ along different directions around a point. The right
thing to do, therefore, is to take the change in the
electromagnetic mass for $|\Delta m|$ given in (\ref{430}) and
(\ref{460})-(\ref{470}) and insert them in (\ref{560}) to see if
the inequalities are violated. The main contribution comes from
the potential due to Earth. The result is $|\Delta m|/m \leq
h\times 10^{-19}$, where $h$ is a length given in meters which is
characteristic of the size of the object measured. In the  case of
an experiment on the Zeeman effect of a nucleus,  $h$ must be
taken as the diameter of a nucleus, {\it i.e.} $h\simeq 10^{-14}$
m, from which we can say that, in any case, the effect of the
variation of $\alpha$ verifies in this model,
 \begin{equation}
{\Delta m\over m} \leq (-\chi +\xi ) \times 10^{-33}. \label{570} \end{equation}
Since $\chi =1$ and $\xi \leq 10^{-5}$  it is seen that the anisotropy in this model is well
below the bounds imposed by the Hughes-Drever experiments. It can
be remarked that in experiments of this kind, the direction of the
magnetic field is usually fixed with respect to the radius from
the center of the Earth. In that case, the main effect would be
due to the Sun which is $10^{3}$ times smaller (\ref{460}).

\section{Conclusions}
The situation of the attenuating quantum vacuum model
\cite{Ran03a,Ran03b,Ran03c} with respect to the experiments has
been studied. This model gives a unified explanation of the
Pioneer anomalous acceleration $a_{\rm P}\simeq 8.5\times
10^{-10}\mbox{ m/s}^2$ observed by Anderson et al \cite{And98} in
four different spacecrafts and of the cosmological variation of
the fine structure constant detected by Webb et al \cite{Web01}.
Its most important prediction is that there must be an adiabatic
universal acceleration of light $a_\ell$, equal to $a_{\rm P}$,
the observational signature of which is the same as that of an
extra constant acceleration directed towards the Sun, {\it i. e.}
a blue shift of the radio signal from the spaceship that increases
linearly in time, indeed just what was observed. The model is
based on an argument that shows that the quantum vacuum must be
decreasing its optical density, what explains why it was dubbed
as ``attenuating quantum vacuum model". Accordingly, the
empty space ({\it i. e.} the quantum vacuum) is treated
phenomenologically as an optical medium, its permittivity and
permeability $\epsilon _0$ and $\mu _0$ being changed
 to $\epsilon _{\rm r}\epsilon _0$ and $\mu _{\rm r}\mu
_0$, where the relative quantities $\epsilon _{\rm r},\, \mu _{\rm
r}$ (of course, equal to 1 in a reference
 terrestrial laboratory) depend on the gravitational potential $\Phi$ through the
expressions (\ref{30}) that include two positive coefficients
$\beta$ and $\gamma$ . It was argued in \cite{Ran03c}
that the best values for them are $\beta \simeq 0.5$, $\gamma
\simeq 1.5$ ( in this paper, the coefficients $\xi
=(3\beta-\gamma)/2$, $\chi =(\beta +\gamma )/2$ are used instead).

As a consequence, the electron charge, the light speed and the
fine structure constant depend on space and time through the
universe as
\begin{eqnarray}
e({\bf r},t)&=& e[1+ {\chi +\xi\over 2} \, {\Phi -\Phi _{\rm R}\over c^2}],\label{580}\\
c({\bf r},t) &=& c[1+\chi \, {\Phi -\Phi _{\rm R}\over c^2}],\label{590}\\
\alpha ({\bf r},t) &=& \alpha [1+ \xi \, {\Phi -\Phi _{\rm R}\over
c^2}], \label{600}
\end{eqnarray}
where $e,\,c,\,\alpha $ are their values at the reference terrestrial
laboratory R at present time and $\Phi _{\rm R}$ is the potential at R.
These equations  fix the values of the coefficients. Einstein
formula (\ref{190}) indicates that $\chi =1$ \cite{Ran04}. The
observations by Webb et al \cite{Web01} give $\xi \simeq 10^{-5}$.
If they have overestimated the variation as some people suggest
today, then $\xi $ would be smaller. If there is no cosmological
variation of $\alpha$ then $\xi =0$ (so that $\beta =0.5$ and
$\gamma =1.5$).

The results of the main experiments concerning special relativity
and the equivalence principle have been considered. The
conclusions can be summarized as follows.

(i) In the Hughes-Drever experiments, the predicted consequences
of the model on the anisotropy of the mass are at least ten orders
of magnitude smaller than the experimental bound. The model fully
agrees, therefore, with these bounds.

(ii) Since the electron charge would be differently renormalized
at different places of the universe, its value would depend on
space and time. As a consequence, the electromagnetic mass of a
body (due to the Coulomb interaction in their nuclei) would vary
 when it moves. A new reaction force equal to
$-\dot{m}{\bf v}$ would act on it, that would depend on its
chemical composition since different nuclei have different
proportions of Coulomb mass. The consequence is that the weak
equivalence principle would be violated, since different bodies
with the same initial conditions would follow different
trajectories in a gravitational field. In principle, this could be observed in
free fall or in E\"otv\"os experiments. However, the variation of $\Phi$
 being so small, this effect would completely unobservable here at
Earth. Indeed, it would be several orders of magnitude too small,
about six orders in the case of the best E\"otv\"os type
experiments by Roll, Krotkov and Dicke \cite{Rol64} or Braginsky
and Panov \cite{Bra72}.

iii) If $\xi =0$ ({\it i. e.} if there is no cosmological
variation of $\alpha$), the model predicts exactly the same result as general relativity
for the gravitational redshift, while it gives exactly half the value for the bending angle of light passing
near the Sun and  for the Shapiro time delay.
This means that it gives exactly the same result as the weak
equivalence principle, which is just a component of general
relativity.

iv) If $\xi \neq 0$, the model predicts a cosmological variation
of $\alpha$ (remember that, if the observations by Webb et al
\cite{Web01} are correct, $\xi\simeq 10^{-5}$). The predictions
concerning the light ray bending around the Sun and the Shapiro
time delay would be the same as for $\xi =0$ (i.e. again the same
as predicted by the weak equivalence principle). There would be,
however, a difference between the predictions of the model for the
gravitational redshift and the observational results, its relative value
being a few times $\xi$. The model agrees with the most stringent
experimental bound by Vessot and Levine if $\xi \lesssim 3\times 10^{-5}$,
while to explain the observations by Webb et al \cite{Web01}, $\xi \simeq 1.3\times
10^{-5}$, just borderline.  However,
there are reasons to believe that Webb et al overestimated the
variation, since some researchers have make calculations
indicating that either the effect is clearly smaller or it is nil
(see \cite{Sri04} for instance). In that case, $\xi$ would be
smaller or even zero, and the prediction of the model would be
well below any experimental bound on the gravitational redshift.

v) Concerning the light acceleration $a_\ell\simeq 8.5\times
10^{-10}\mbox{ m/s}^2$ predicted by this model, it must be
emphasized that it is very small. During a year, the light speed
would increase in about $2.7$ cm/s, just about 1 part in
$10^{10}$. In 1983, the light velocity was fixed at the measured
value of the time, which has an error margin of about 1 m/s. Since
that moment the change would have been about 56 cm/s. If such a
variation can be detected in an experiment, the change in $c$
since 1983 could be measured. It was stated in section 5.1 that
 a complete analysis of this acceleration requires to consider a
 cosmological time $t_{\rm cosm}$ which is accelerated with
 respect to the parametric time $t$, the acceleration being the Hubble constant.

vi) The fact that the light propagation depends on the metric
indicates that a gravitational field has an effect on the Maxwell
equations, a fact first recognized by Einstein in 1907. Since the effect
is small in weak gravity, it seems clear that the problem can be
treated by introducing a dependence on the gravitational potential
$\Phi$ of the permittivity and permeability of empty space, as done
in \cite{Ran03a,Ran03b,Ran03c}. It was
shown  in \cite{Ran04} and in section 4  that the
existence of an acceleration of light $a_\ell$ can be deduced from a
general relativistic argument based on Einstein formula (\ref{190}).
Furthermore, it was shown in section 2 that an estimation gives a
value close to $a_{\rm P}$. A way to characterize then the situation
 could be the following: the attenuating vacuum
model here studied gives an alternative (and equivalent) description to the relation between
gravity and electromagnetism, first recognized by Einstein in 1907,
when proposing his equivalence principle.
This implies that the Newtonian model
proposed in \cite{Ran03a,Ran03b,Ran03c} can be integrated in
general relativity, provided that $\chi =1$. If $\xi =0$, it is
certain that there would be no problem for that. In that case, the
 effect of gravitation on Maxwell equations could be interpreted to
 be due to its effect on the quantum vacuum, the optical density of which
would depend on the potential $\Phi$ as this model proposes. It
must be kept in mind that the quantum vacuum fixes the observed
values of some important quantities by means of a process of
renormalization.

vii) If $\xi$ does not vanish but is small enough, say $|\xi|\leq
10^{-5}$, the integration of the model in general relativity would
still be possible. The effect of the attenuation of the quantum
vacuum could be observed as a cosmological evolution of $\alpha$,
probably smaller than what was reported (with the advantage that
it could be compatible the with the bound obtained from the Oklo
reactor). Indeed, other experiments are needed to determine the
right value of the variation of $\alpha$ over cosmological
intervals of time, or even if it varies at all.

viii) The fact that the model can be integrated in general relativity,
just giving another interpretation of Einstein formula (\ref{190})
fully compatible with the rest, has an important consequence. If $\xi =0$
the trajectories of light would be the same as in general relativity,
for low curvature at least.
This would imply that the Global Positioning System would work just as
predicted, so that the variation of the light speed could not be detected
from its results. If $\xi <10^{-5}$ the effect would be still too small to affect
the positions given by it. The same would apply to other experiments on light.

To summarize, the attenuating quantum vacuum model must be further
studied since, being in agreement with the experimental data, it
could give a good insight of some important phenomena and could
have interesting cosmological implications. The adiabatic
acceleration of light is an example. In any case, the explanation
of the Pioneer anomalous acceleration proposed by this model must
be studied by the experts, in particular by those who have the
data of the trajectories of the ships.

\section{Acknowledgements}
I am very indebted to Profs. A. Tiemblo and J.L. Sebasti\'an for discussions.

\end{document}